%% file: main.tex
\documentclass[%
 reprint,
 superscriptaddress,
 amsmath,amssymb,
 aps,
 prl,
]{revtex4-2}

\usepackage{orcidlink}
\usepackage{graphicx}% Include figure files
\usepackage{dcolumn}% Align table columns on decimal point
\usepackage{bm}% bold math
\usepackage{xcolor}
\usepackage[mathlines]{lineno}% Enable numbering of text and display math
\begin{document}

\preprint{APS/123-QED}

\title{Measurements of Angular Distributions of Drell-Yan Dimuons in $p+p$ and $p+d$ Interactions at 120 GeV/$c$}% Force line breaks with \\
%\thanks{aaa}%

\input{author-list}

\date{\today}% It is always \today, today,

\begin{abstract}
\input{abst}  %%% KEI
\end{abstract}

\maketitle

\input{intro} %%%%%%% Theoretical background Kenichi

\input{status}

\input{e906} %%%%%% SeaQuest setup/analysis   %%% Kei
\input{res_discuss} %%%%%%%% Results and discussion   %%% Kei
%\nocite{*}
\bibliography{ref}% Produces the bibliography via BibTeX.

\end{document}

%% file: author-list.tex
\author{K.~Nagai}
\email{knagai@memphis.edu}
%\orcid{0000-0002-5336-8306}
\affiliation{
    Department of Physics and Materials Science, The University of Memphis, Memphis, TN 38152
}%
\affiliation{
    Los Alamos National Laboratory, Los Alamos, New Mexico 87545, US
}
\affiliation{
    Institute of Physics, Academia Sinica, Taipei,11529, Taiwan
}
\affiliation{
    Tokyo Institute of Technology, Tokyo, Japan
}

\author{M.F.~Hossain}
%\orcid{0000-0002-6467-1394}
\affiliation{
    New Mexico State University, Las Cruces, NM 88003 USA
}
\affiliation{
    University of Virginia, Charlottesville, VA, 22904 USA 
}

\author{A.~Pun}
\affiliation{
    New Mexico State University, Las Cruces, NM 88003 USA
}

\author{K.~Nakano}%
%\orcid{0000-0002-8925-2233}
\affiliation{
    University of Virginia, Charlottesville, VA, 22904 USA 
}
\affiliation{
    Tokyo Institute of Technology, Tokyo, Japan
}
\affiliation{
    RIKEN Nishina Center for Accelerator-Based Science, Wako, Saitama 351-0198, Japa
}

\author{K.~Liu}%
%\orcid{0000-0002-6676-8165}
\affiliation{
    Los Alamos National Laboratory, Los Alamos, New Mexico 87545, US
}

\author{C.~A.~Aidala}
%\orcid{0000-0001-9540-4988}
\affiliation{
    University of Michigan, Ann Arbor, Michigan 48109, USA
}
\affiliation{
    Los Alamos National Laboratory, Los Alamos, New Mexico 87545, US
}

\author{J.~Arrington}
%\orcid{0000-0002-0702-1328}
\altaffiliation[Present Address:]{
    Lawrence Berkeley National Laboratory
}
\affiliation{
    Physics Division, Argonne National Laboratory, Lemont, Illinois 60439, USA
}

\author{C.~Ayuso}
\affiliation{
    University of Michigan, Ann Arbor, Michigan 48109, USA
}

\author{C.~Barker}
\affiliation{
    Abilene Christian University, Abilene, Texas 79699 USA
}

\author{W.C.~Chang}
%\orcid{0000-0002-1695-7830}
\affiliation{
    Institute of Physics, Academia Sinica, Taipei,11529, Taiwan
}

\author{A.~Chen}
\affiliation{
    University of Illinois at Urbana-Champaign, Urbana, Illinois 61801, USA
}

\author{D.C.~Christian}
%\orcid{0000-0003-1275-6510}
\affiliation{
    Fermi National Accelerator Laboratory, Batavia, Illinois 60510, USA
}

\author{B.~Dannowitz}
\affiliation{
    University of Illinois at Urbana-Champaign, Urbana, Illinois 61801, USA
}

\author{J.~Dove}
\affiliation{
    University of Illinois at Urbana-Champaign, Urbana, Illinois 61801, USA
}

\author{L.El~Fassi}
%\orcid{0000-0003-3647-3136}
\affiliation{
    Mississippi State University, Mississippi State, Mississippi 39762, USA
}
\affiliation{
    Rutgers, The State University of New Jersey, Piscataway, New Jersey 08854, USA
}

\author{D.F.~Geesaman}
%\orcid{0000-0003-2557-3131}
\affiliation{
    Physics Division, Argonne National Laboratory, Lemont, Illinois 60439, USA
}

\author{R.~Gilman}
\affiliation{
    Rutgers, The State University of New Jersey, Piscataway, New Jersey 08854, USA
}

\author{Y.~Goto}
%\orcid{0000-0002-2973-7458}
\affiliation{
    RIKEN Nishina Center for Accelerator-Based Science, Wako, Saitama 351-0198, Japa
}

\author{R.S.~Guo}
\affiliation{
    Department of Physics, National Kaohsiung Normal University, Kaohsiung 824, Taiwan
}

\author{L.~Guo}
\altaffiliation[Present Address:]{
    Physics Department, Florida International University
}
\affiliation{
    Los Alamos National Laboratory, Los Alamos, New Mexico 87545, US
}

\author{T.J.~Hague}
\altaffiliation[Present Address:]{
    Lawrence Berkeley National Laboratory
}
\affiliation{
    Abilene Christian University, Abilene, Texas 79699 USA
}

\author{R.J.~Holt}
%\orcid{0000-0001-9225-9914}
\altaffiliation[Present Address:]{
    Kellogg Radiation Laboratory, California Institute of Technology, Pasadena, California 91125
}
\affiliation{
    Physics Division, Argonne National Laboratory, Lemont, Illinois 60439, USA
}

\author{D.~Isenhower}
%\orcid{0000-0002-8237-5636}
\affiliation{
    Abilene Christian University, Abilene, Texas 79699 USA
}

\author{E.R.~Kinney}
%\orcid{0000-0002-4176-5283}
\affiliation{
    University of Colorado, Boulder, Colorado 80309, USA
}

\author{A.~Klein}
\affiliation{
    Los Alamos National Laboratory, Los Alamos, New Mexico 87545, US
}

\author{D.~Kleinjan}
\affiliation{
    Los Alamos National Laboratory, Los Alamos, New Mexico 87545, US
}

%\author{C.D.~Kuruppu}
%\affiliation{
%    New Mexico State University, Las Cruces, NM 88003 USA
%}

\author{C.H.~Leung}
%\orcid{0000-0001-7907-3728}
\altaffiliation[Present Address:]{
    Thomas Jefferson National Accelerator Facility, Newport News, US 
}
\affiliation{
    University of Illinois at Urbana-Champaign, Urbana, Illinois 61801, USA
}

\author{P.-J.~Lin}
%\orcid{0000-0001-7073-6839}
\altaffiliation[Present Address:]{
    Department of Physics, National Central University, Jhongli District, Taoyuan City 32001,Taiwan
}
\affiliation{
    University of Colorado, Boulder, Colorado 80309, USA
}

\author{M.X.~Liu}
\affiliation{
    Los Alamos National Laboratory, Los Alamos, New Mexico 87545, US
}

\author{W.~Lorenzon}
%\orcid{0000-0003-0657-8463}
\affiliation{
    University of Michigan, Ann Arbor, Michigan 48109, USA
}

\author{R.E.~McClellan}
\altaffiliation[Present Address:]{
    Pensacola State College, Pensacola, FL 32504
}
\affiliation{
    University of Illinois at Urbana-Champaign, Urbana, Illinois 61801, USA
}

\author{P.L.~McGaughey}
\affiliation{
    Los Alamos National Laboratory, Los Alamos, New Mexico 87545, US
}

\author{M.M.~Medeiros}
\affiliation{
    Physics Division, Argonne National Laboratory, Lemont, Illinois 60439, USA
}

\author{Y.~Miyachi}
%\orcid{0000-0002-8502-3177}
\affiliation{
    Yamagata University, Yamagata, Japan
}

\author{S.~Miyasaka}
\affiliation{
    Tokyo Institute of Technology, Tokyo, Japan
}

\author{D.H.~Morton}
\affiliation{
    University of Michigan, Ann Arbor, Michigan 48109, USA
}

\author{K.~Nakahara}

\affiliation{
    University of Maryland, College Park, Maryland 20742, USA
}

\author{J.C.~Peng}
\affiliation{
    University of Illinois at Urbana-Champaign, Urbana, Illinois 61801, USA
}

\author{S.~Prasad}
\affiliation{
    University of Illinois at Urbana-Champaign, Urbana, Illinois 61801, USA
}
\affiliation{
    Physics Division, Argonne National Laboratory, Lemont, Illinois 60439, USA
}

\author{A.J.R.~Puckett}
\altaffiliation[Present Address:]{
    University of Connecticut, Storrs, CT, 06269, USA
}
\affiliation{
    Los Alamos National Laboratory, Los Alamos, New Mexico 87545, US
}

\author{B.J.~Ramson}
\affiliation{
    University of Michigan, Ann Arbor, Michigan 48109, USA
}
\affiliation{
    Fermi National Accelerator Laboratory, Batavia, Illinois 60510, USA
}

\author{P.E.~Reimer}
%\orcid{0000-0002-0301-2176}
\affiliation{
    Physics Division, Argonne National Laboratory, Lemont, Illinois 60439, USA
}

\author{J.G.~Rubin}
\affiliation{
    University of Michigan, Ann Arbor, Michigan 48109, USA
}
\affiliation{
    Physics Division, Argonne National Laboratory, Lemont, Illinois 60439, USA
}

\author{F.~Sanftl}
\affiliation{
    Tokyo Institute of Technology, Tokyo, Japan
}

\author{S.~Sawada}
\affiliation{
    KEK, High Energy Accelerator Research Organization, Tsukuba, Ibaraki 305-0801, Japan
}

\author{T.~Sawada}
\altaffiliation[Present Address:]{
    Institute for Cosmic Ray Research, The University of Tokyo, Hida, Gifu 506-1205, Japan
}
\affiliation{
    University of Michigan, Ann Arbor, Michigan 48109, USA
}

\author{M.B.C.~Scott}
\altaffiliation[Present Address:]{
    George Washington University, Washington, DC 20052, USA
}
\affiliation{
    University of Michigan, Ann Arbor, Michigan 48109, USA
}
\affiliation{
    Physics Division, Argonne National Laboratory, Lemont, Illinois 60439, USA
}

\author{T.-A.~Shibata}
%\orcid{0009-0005-5498-4804}
\altaffiliation[Present Address:]{
    Nihon University, College of Science and Technology, Chiyoda-ku, Tokyo 101-8308, Japan
}
\affiliation{
    Tokyo Institute of Technology, Tokyo, Japan
}
\affiliation{
    RIKEN Nishina Center for Accelerator-Based Science, Wako, Saitama 351-0198, Japa
}

\author{D.-S.~Su}
\affiliation{
    Institute of Physics, Academia Sinica, Taipei,11529, Taiwan
}

\author{A.S.~Tadepalli}
\altaffiliation[Present Address:]{
    Thomas Jefferson National Accelerator Facility, Newport News, US 
}
\affiliation{
    Rutgers, The State University of New Jersey, Piscataway, New Jersey 08854, USA
}

\author{B.G.~Tice}
\affiliation{
    Physics Division, Argonne National Laboratory, Lemont, Illinois 60439, USA
}

\author{R.S.~Towell}
%\orcid{0000-0003-3640-7008}
\affiliation{
    Abilene Christian University, Abilene, Texas 79699 USA
}

\author{S.~Uemura}
%\orcid{0000-0003-3458-4625}
\altaffiliation[Present Address:]{
    Fermi National Accelerator Laboratory, Batavia, Illinois 60510, USA
}
\affiliation{
    Los Alamos National Laboratory, Los Alamos, New Mexico 87545, US
}

\author{S.G.~Wang}
\altaffiliation[Present Address:]{
    Advanced Photon Source, Argonne National Laboratory, Lemont, Illinois 60439, USA
}
\affiliation{
    Institute of Physics, Academia Sinica, Taipei,11529, Taiwan
}
\affiliation{
    Department of Physics, National Kaohsiung Normal University, Kaohsiung 824, Taiwan
}
\altaffiliation[Present Address:]{
    Fermi National Accelerator Laboratory, Batavia, Illinois 60510, USA
}

\author{J.~Wu}
\affiliation{
    Fermi National Accelerator Laboratory, Batavia, Illinois 60510, USA
}

\author{N.~Wuerfel}
\affiliation{
    University of Michigan, Ann Arbor, Michigan 48109, USA
}

\author{Z.H.~Ye}
\altaffiliation[Present Address:]{
    Department of Physics, Tsinghua University, Beijing China 
}
\affiliation{
    Physics Division, Argonne National Laboratory, Lemont, Illinois 60439, USA
}

\collaboration{FNAL E906/SeaQuest Collaboration}\noaffiliation

%% file: abst.tex
%We present a detailed analysis of the angular distributions of Drell–Yan dimuons produced by a 120 GeV/c proton beam interacting with liquid hydrogen and deuterium targets. 
We present experimental results on the angular distributions of Drell-Yan muons produced by a 120 GeV/$c$ proton beam interacting with liquid hydrogen and deuterium targets. 
The dimuon angular distributions in both polar ($\theta$) and azimuthal ($\phi$) angles in the Collins-Soper frame are measured within the kinematic range of $4.5 < m_{\mu\mu} < 10\ \mathrm{GeV}/c^2$, $0.19 < p_T < 2.24\ \mathrm{GeV}/c$, and $0 < x_F < 0.95$. 
Unlike the results of a previous proton-induced Drell-Yan experiment at a higher energy, %a pronounced $\cos 2\phi$ modulation is observed in our proton-induced Drell–Yan data. 
the data reveal a pronounced $\cos 2\phi$ modulation in the angular distributions.
Comparison with perturbative QCD (pQCD) predictions shows statistically significant deviations, with p-values of 3.5\% for the $p+p$ and 1.5\% for the $p+d$ Drell-Yan processes. 
These results suggest the presence of nonperturbative QCD contributions.

%%%%%%% inconsistent with E866

%% file: intro.tex
\begin{figure}[hbtp] \centering
    \includegraphics[width=\linewidth]{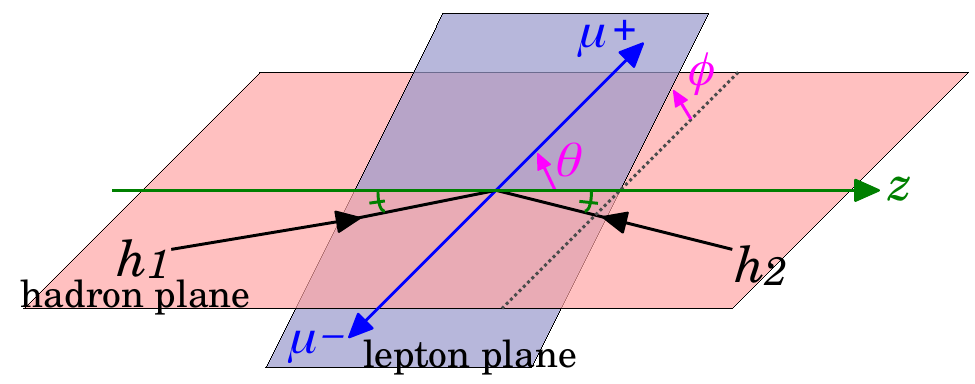}
    \caption{Angle definitions of the Drell-Yan process in the Collins-Soper frame.
        It is a rest frame of the virtual photon.
        The hadron plane is defined by the momenta of two interacting hadrons ($h_1$ and $h_2$).
        The $z$ axis is taken on the hadron plane so that the angle to $h_1$ is equal to that to $h_2$.
        The lepton plane is defined by the momenta of two final-state leptons ($\mu^+$ and $\mu^-$) and the $z$ axis.}
    \label{fig:collins_soper_frame}
\end{figure}

%The Drell--Yan process has played an important role in understanding high energy hadron+hadron interactions since it was discovered in 1970~\cite{Drell:1970wh,Christenson:1970um}.
The Drell-Yan process was proposed in 1970~\cite{Drell:1970wh,Christenson:1970um} to describe the underlying mechanism for high-mass dimuon production in hadron-hadron interactions.
In this process a quark and an antiquark in two colliding hadrons annihilate into a virtual photon that then decays into a lepton-antilepton pair.
It is one of the cleanest processes in high energy %hadron-hadron reactions because it does not undergo any hadronic final-state interactions with only leptons in the final state.
hadron-hadron collisions because the final state leptons have no QCD final-state interactions.
Therefore it can be used to directly access the scattering process of partons and the internal structure of scattering hadrons, with no correction for final state interaction involved.
For the proton-induced Drell-Yan processes, the dominant contribution comes from the $u\bar{u}$ annihilation, where $u$ and $\bar{u}$ originate from the proton beam and target nucleons, respectively.
The $u\bar{u}$ channel dominates over the $d\bar{d}$ channel because the Drell-Yan cross section is proportional to the square of the quark's electric charge.
%The Drell--Yan process takes place via $u\bar{u}$ or $d\bar{d}$ annihilation. The $u\bar{u}$ channel is dominant because of the electric charge squared. % Shibata-san's comment
%Particularly 
The angular distribution of lepton pairs from the Drell-Yan process is 
%of great interest, as it is 
sensitive to transverse-momentum-dependent parton distribution functions (TMD PDFs)~\cite{Boer:1997nt}.
It can be expressed in any rest frame of the virtual photon, such as the Collins-Soper frame~\cite{Collins:1977iv}, as:
\begin{linenomath}\begin{multline}
  \frac{1}{\sigma}\frac{d\sigma}{d\Omega} =\frac{3}{4\pi} \frac{1}{\lambda + 3} \biggl( 1 + \lambda \cos^2\theta \\
    + \mu \sin2\theta \cos\phi + \frac{\nu}{2} \sin^2\theta \cos2\phi\biggr), \label{eq:cross-section}
\end{multline} \end{linenomath}
where $\phi$ and $\theta$ are the azimuthal and polar angles, 
as illustrated in Fig.~\ref{fig:collins_soper_frame}, and
$\lambda$, $\mu$ and $\nu$ represent the magnitude of the angular dependence.
%The angular distribution of the Drell--Yan process has so far been measured with various types of beams and targets in order to disentangle two contributions that determine the angular distribution, as explained below.
%multiple factors that determine the angular distribution.
%the various terms in Eq.~(\ref{eq:cross-section}), as explained later.

The cross section of the Drell-Yan process can be computed based on perturbative Quantum Chromodynamics (pQCD)~\cite{Collins:1989gx}.
Ignoring intrinsic transverse momentum of the quarks, at leading order (LO), the transverse momentum ($p_T$) of the virtual photon is zero, and the angular dependence becomes $1 + \cos^2\theta$ with $\lambda = 1$, $\mu = 0$ and $\nu = 0$.
At higher order in $\alpha_s$, the processes such as gluon emission would lead to nonzero transverse momentum ($p_T$) of the virtual photon, and thus to a nonzero $\phi$ dependence, i.e.~$\mu \ne 0$ and $\nu \ne 0$.
%$k_T$ becomes nonzero, which leads to nonzero transverse momentum ($p_T$) of the virtual photon and then nonzero $\phi$ dependence, i.e.~$\mu \ne 0$ and $\nu \ne 0$.
%It is well known as the Lam--Tung relation~\cite{Lam:1978pu} that a relation of $1 - \lambda = 2\nu$ is preserved at the next-to-leading order (NLO; $\alpha_s$) but not at the next-to-next-to-leading order (NNLO; $\alpha_s^2$).
%The well known Lam--Tung relation~\cite{Lam:1978pu} states that a relation of $1 - \lambda = 2\nu$ is preserved at the next-to-leading order (NLO; $\alpha_s$) but not at the next-to-next-to-leading order (NNLO; $\alpha_s^2$)~\cite{PhysRevD.21.2712,PhysRevD.51.4891}.
%Therefore the study of the angular coefficients and the Lam--Tung relation helps us to examine the higher-order effects.
Theoretical calculations at next-to-next-to-leading order (NNLO; $\alpha_s^2$) have been performed and compared to existing experimental datasets~\cite{PhysRevD.93.114013,PhysRevD.99.014032}.
The calculations generally describe well the observed trends of $\lambda$, $\mu$ and $\nu$ as functions of the transverse momentum ($p_T$) of the virtual photon, but exhibit significant deviations in several experimental datasets and $p_T$ ranges.

%The angular distributions also reflect the state of interacting partons.
There is another mechanism that affects the angular distributions.
The Boer-Mulders function ($h_1^\perp(x, k_T)$)~\cite{Boer:1997nt} is a TMD PDF that represents the correlation between the transverse spin and the transverse momentum of parton within an unpolarized hadron.
The coefficient $\nu$ is roughly proportional to the product of the Boer-Mulders functions of two interacting partons; 
$\nu \propto h_1^{\perp f_1}(x_1, k_{T1}) \times h_1^{\perp f_2}(x_2, k_{T2})$~\cite{PhysRevD.60.014012},
where $f_1$, $x_1$ and $k_{T1}$ denote the flavor, the Bjorken $x$ and the transverse momentum of the interacting parton in the beam hadron, and $f_2$, $x_2$ and $k_{T2}$ in the target hadron.
Therefore $h_1^\perp(x, k_T)$ can be constrained by a measurement of $\nu$.
Extractions of $h_1^\perp(x, k_T)$ from existing experimental datasets have been investigated, although the statistics of these datasets are highly limited%at present
~\cite{Zhang:2008nu,Lu:2009ip,Barone:2010gk,Barone:2009hw,Wang:2018naw,Christova:2020ahe,Piloneta:2024aac}.%\cite{Zhang:2008nu,Wang:2018naw}.
%Experiments of semi-inclusive deep-inelastic scattering (SIDIS) discovered that
%$h_1^\perp(x, k_T)$ of quarks is non-zero\cite{Barone:2009hw}.
%Also $h_1^\perp(x, k_T)$ of anti-quarks was extracted
%from the Drell-Yan process measured by the E866/NuSea experiment\cite{Zhu:2006gx,Zhu:2008sj},
%although the statistical accuracy is quite limited\cite{Lu:2009ip,Barone:2010gk}.
%SeaQuest aims to measure $h_1^\perp(x, k_T)$ of anti-quarks with a better precision with the Drell-Yan process.

Both mechanisms described above---higher-order pQCD effects and the Boer-Mulders function---may contribute to the observed angular distribution. 
The relative contributions of these two effects can be disentangled through systematic measurements using different beam and target combinations.

%% file: status.tex
The NA10 and E615 collaborations conducted early pion-induced Drell-Yan studies, measuring the angular distributions in $\pi^-N$ collisions. 
The NA10 data \cite{NA10_2}, using $\pi^-$ beams at different energies on deuterium and tungsten targets, showed no dependence on center-of-mass energy or nuclear effects.
The $\lambda$ parameter was close to unity, consistent with the naive parton model, and $\mu$ was near zero, indicating balanced transverse momentum contributions from annihilating partons. 
However, the most striking result was the large $\nu$ parameter, which reached values as high as 0.3 and increased strongly with the dimuon transverse momentum $p_T$, a behavior inconsistent with pQCD predictions~\cite{PhysRevD.99.014032}.
The E615 experiment at Fermilab \cite{E615_1,E615_2} confirmed these results with a broader kinematic range in $\pi^-N\rightarrow\mu^+\mu^-X$, finding similar large $\nu$ values and $p_T$ dependence. 
Both NA10 and E615 observed significant $\mathrm{cos}(2\phi)$
angular asymmetries, indicating the presence of transverse momentum dependent effects beyond the collinear factorization framework~\cite{PhysRevD.99.014032}. 
These findings were interpreted as evidence of the Boer-Mulders mechanism, which introduces spin-momentum correlations of partons.
Proton-induced Drell-Yan studies by the E866/NuSea collaboration at Fermilab revealed a contrasting picture. 
In both $p+p$ \cite{E866pp} and $p+d$ \cite{E866pd} collisions at 800 GeV/$c$, much smaller $\nu$ values ($<0.05$) were observed. A possible interpretation for this dramatic difference was that it likely arises from the underlying partonic structure: while pion-induced processes involve valence antiquarks from the pion beam annihilating with valence quarks from the nucleon target, proton-induced processes primarily access sea antiquarks from the target nucleon. 
%The results suggest 
This suggests that Boer-Mulders functions for sea antiquarks are significantly smaller than those for valence quarks, providing crucial constraints on the flavor dependence of transverse momentum dependent parton distributions.

 At collider energies, studies of lepton angular distributions in $W$ and $Z$ boson production provide complementary perspectives across different kinematic regimes and QCD dynamics. The CDF collaboration \cite{cdfresult} reported the first measurement of lepton angular distributions in the $Z$ mass region with $p+\bar{p}$ collisions, and found good agreement with leading-order pQCD predictions at moderate $p_T$, in contrast to the pion-induced fixed target experiments. Conversely, recent high statistics results in $p+p$ collisions from CMS \cite{cmsresult}, ATLAS \cite{atlasresult} and LHCb \cite{lhcbresult} show clear deviations from naive parton model expectations at high $p_T$ (above 5--10 GeV/$c$), extending up to transverse momenta of 300 GeV/$c$, driven by perturbative NNLO QCD effects rather than intrinsic transverse momentum or Boer-Mulders functions \cite{lhcpqcd}.
 The LHCb results additionally show a significant nonzero and positive $\cos 2\phi$ modulation in a single low-$p_T$ bin ($p_T < 3.5\ \mathrm{GeV}/c$), which may be due to nonperturbative effects.

Overall, measuring angular distributions in both $p+p$ and $p+d$ collisions across energies is crucial for understanding the interplay between sea and valence quark dynamics, as well as perturbative and nonperturbative QCD effects. Specifically, these measurements are vital for the separation of competing contributions to the nonzero $\nu$ parameter from pQCD effects versus Boer-Mulders processes, as the two mechanisms  predict different energy scaling and correlations with spin asymmetries \cite{PhysRevD.99.014032,PhysRevD.60.014012}.
Comparing these results with pion-induced experiments, where interactions between the pion's valence antiquarks with the proton's valence quarks dominate, reveals how different projectiles and energy scales probe unique aspects of QCD, offering a broader perspective on hadronic structure.

%% file: e906.tex
The E906/SeaQuest experiment at Fermilab measured the angular distributions of the proton-induced Drell--Yan process.
The details of the spectrometer are described in Ref. \cite{e906NIM}.
The 120 GeV proton beam provided by the Fermilab Main Injector was incident with the 50.8 cm long liquid hydrogen (LH$_2$) and liquid deuterium (LD$_2$) targets.
A focusing magnet (FMag) positioned immediately downstream of the targets selects muons in the desired momentum range.
The iron magnet also serves as a beam dump, stopping unscattered beam protons from reaching the tracking detectors.
Four tracking stations detected the generated muons.
Each station consisted of hodoscope arrays and drift chambers (Sts.~1--3) or proportional tubes (St.~4).
An open-aperture magnet was placed between the first and second tracking stations to measure the momentum of the muons.
A hadron absorber was placed between the third and fourth tracking stations for muon identification.
The kinematic variables $m_{\mu\mu},\ p_T$ and $x_F$ were reconstructed based on the measured momenta of the muons.
A dimuon mass cutoff of $m_{\mu\mu}>4.5$ GeV/$c^2$ was applied to remove the background from the contribution of charmonium states.
In 2015, $5.6 \times 10^{17}$ protons were delivered to liquid hydrogen and deuterium targets for physics data collection. 
The dataset, which includes approximately 13,000 dimuons from the liquid hydrogen target and 15,000 from the deuterium target within the range $-0.375 < \cos\theta < 0.375$, has been analyzed to extract the Drell–Yan angular distribution coefficients.
%Figure~\ref{fig:e906_e866_pt} and Table~\ref{tab:results_LD2} present the angular distribution parameters $\mu$ and $\nu$ as functions of $p_T$, extracted from the LH$_2$ and LD$_2$ data. 
The Drell-Yan data were divided into three $p_T$ bins %to analyze the $p_T$ dependence of these parameters, 
with approximately equal statistics in each bin. 

Two types of background were identified: %after the initial event selection
dimuon pairs originating from the target flask of the liquid targets, and combinatorial background. Dedicated data were collected with the beam incident on an empty flask to study the first type of background. These empty flask data were normalized by the total number of protons on target and subtracted from the LH$_2$ (LD$_2$) data. 
The combinatorial background, composed of uncorrelated muon pairs, was estimated using the event mixing method~\cite{NMSU_background} applied to the corresponding liquid target data.
After subtracting both types of background, the remaining data were corrected for detector acceptance and detector efficiency effects using Monte Carlo (MC) simulations. 
The MC simulations were weighted to match the measured kinematic distributions of the data. 
These weight parameters were determined by fitting the ratios of simulated data to measured data.

An unbinned-log-likelihood fitting method was used to extract the parameters, with the $\lambda$ value fixed at a certain value
%based on pQCD calculations~\cite{PhysRevD.99.014032} 
due to the limited sensitivity of our spectrometer to $\lambda$.
%The $\mu$ and $\nu$ values were unchanged even when the $\lambda $ was left unfixed.
Since the acceptance and efficiency corrections depend on the $\lambda$, $\mu$, and $\nu$ parameters used in the simulation, an iterative analysis was performed. The data fitting was repeated with simulations updated using angular parameters obtained from the previous fit, %until the differences between simulation and results became negligible.
until the angular parameters converged.

\begin{table}
    \centering
    %\tiny
    \begin{tabular}{c|cc|cc|cc}
    \hline
    \hline
    \multicolumn{7}{c}{LH$_2$}\\
    \hline
    $p_T$ bin (GeV/$c$)& \multicolumn{2}{c|}{0.19--0.55} & \multicolumn{2}{c|}{0.55--0.88} & \multicolumn{2}{c}{0.88--2.24}\\
    Source & $\mu$ & $\nu$ & $\mu$ & $\nu$ & $\mu$ & $\nu$  \\
    \hline
    Acceptance Corr.     & 0.01 & 0.00 & 0.01 & 0.01 & 0.03 & 0.01 \\
%    Rate Corr. Err.     & 0.00 & 0.00 & 0.00 & 0.00 & 0.00 & 0.00 \\
%    Rate Corr. Func.    & 0.00 & 0.00 & 0.00 & 0.00 & 0.01 & 0.00 \\
%    Protons on Target   & 0.00 & 0.00 & 0.00 & 0.00 & 0.01 & 0.00 \\
    Kinematic Weight     & 0.01 & 0.00 & 0.02 & 0.01 & 0.04 & 0.02 \\
    Resolution          & 0.02 & 0.03 & 0.02 & 0.02 & 0.02 & 0.02 \\
    Mass cutoff         & 0.11 & 0.04 & 0.03 & 0.01 & 0.03 & 0.02 \\
    Non-fixed $\lambda$ & 0.01 & 0.00 & 0.02 & 0.00 & 0.00 & 0.01 \\
    Others              & 0.00 & 0.01 & 0.00 & 0.00 & 0.01 & 0.00 \\
    \hline
    Systematic Total    & 0.11 & 0.05 & 0.05 & 0.03 & 0.07 & 0.04   \\
    \hline 
    Statistical Err. & 0.1 & 0.06 & 0.1 & 0.09 & 0.2 & 0.1 \\
    \hline
%    \end{tabular}
%    \caption{Breakdown of systematic uncertainties for LH$_2$.}
%    \label{tab:syst_LH2}
%\end{table}
%\begin{table}[htbp]
%    \centering
%    %\tiny
%    \begin{tabular}{c|cc|cc|cc}
    \hline
    \multicolumn{7}{c}{LD$_2$}\\
    \hline
    $p_T$ bin (GeV/$c$Ge)& \multicolumn{2}{c|}{0.19-0.55} & \multicolumn{2}{c|}{0.55-0.88} & \multicolumn{2}{c}{0.88-2.24}\\
%    & \multicolumn{2}{c|}{1st $p_T$ bin} & \multicolumn{2}{c|}{2nd $p_T$ bin} & \multicolumn{2}{c}{3rd $p_T$ bin}\\
    Source & $\mu$ & $\nu$ & $\mu$ & $\nu$ & $\mu$ & $\nu$  \\
    \hline
    Acceptance Corr.      & 0.01 & 0.00 & 0.01 & 0.00 & 0.03 & 0.01 \\
%    Rate Corr. Err.     & 0.00 & 0.00 & 0.00 & 0.00 & 0.00 & 0.00 \\
%    Rate Corr Func.     & 0.00 & 0.00 & 0.00 & 0.00 & 0.01 & 0.01 \\
%    Protons on Target   & 0.00 & 0.00 & 0.00 & 0.00 & 0.00 & 0.00 \\
    Kinematic Weight   & 0.01 & 0.01 & 0.02 & 0.01 & 0.06 & 0.03 \\ 
    Resolution          & 0.01 & 0.00 & 0.01 & 0.01 & 0.03 & 0.01 \\
    Mass Cutoff         & 0.02 & 0.03 & 0.05 & 0.03 & 0.03 & 0.00 \\
    Non-fixed $\lambda$ & 0.00 & 0.00 & 0.00 & 0.00 & 0.00 & 0.00 \\
    LD$_2$ Purity       & 0.02 & 0.01 & 0.03 & 0.00 & 0.01 & 0.00 \\
    Others              & 0.00 & 0.00 & 0.00 & 0.00 & 0.01 & 0.01 \\
    \hline
    Systematic Total    & 0.03 & 0.04 & 0.06 & 0.04 & 0.08 & 0.03  \\
    \hline
    Statistical Err. & 0.08 & 0.06 & 0.10 & 0.06 & 0.2 & 0.08\\
    \hline
    \hline
    \end{tabular}
    \caption{Systematic uncertainties for LH$_2$ target (top) and LD$_2$ target (bottom) events. The ``Others'' row includes the combined systematic uncertainties from the rate dependent efficiency corrections and the uncertainty from the beam normalization. }%the number of the total protons on target.}%, and $\lambda$ value fixed at pQCD calculations.}
    \label{tab:syst_LD2}
\end{table}

Table~\ref{tab:syst_LD2} summarizes the sources and values of systematic uncertainties for the LH$_2$ and LD$_2$ data. 
The most significant sources are the acceptance correction, kinematic weighting, detector resolution, and the dimuon minimum mass. 
The uncertainty from the acceptance correction was estimated using MC simulations and propagated to the measured angular distribution parameters. 
The uncertainties of the parameters from the additional weight on the MC simulations were also propagated to estimate their impact on the final results.
%An additional weight was applied to the MC simulations to align the generated kinematic distributions with the measured data. 
%These their uncertainties were propagated to estimate their impact on the final results.
The uncertainty from the detector resolution was evaluated by varying $\cos\theta$ and $\phi$ by $\pm10\%$ and comparing the fit results. 
For the dimuon mass cutoff, a threshold of 4.5 GeV/$c^2$ was used, and systematic uncertainties were assessed by performing analyses with cutoff masses of 4.4 GeV/$c^2$ and 4.6 GeV/$c^2$. 
%The systematic uncertainty due to the LD$_2$ purity was estimated by linear combinations of the LH$_2$ and LD$_2$ results.
The systematic uncertainty due to LD$_2$ target purity was estimated by linear combinations of the LH$_2$ and LD$_2$ results, with the LH$_2$ results representing the hydrogen contamination component.
The $\lambda$ values were fixed based on the pQCD calculations~\cite{PhysRevD.99.014032} in this analysis.
To assess the impact of fixing the $\lambda$ parameter, additional analyses were performed with $\lambda$ allowed to float. 
Although the resulting $\lambda$ values differed by 0.3 to 1.9 compared to the fixed values, the corresponding shifts in the extracted $\mu$ and $\nu$ parameters were minor as shown in Tab.~\ref{tab:syst_LD2}.
These differences were accounted for in the evaluation of the systematic uncertainties.
Additional sources of systematic uncertainty, including rate-dependent efficiency corrections and the beam normalization
%the total number of protons on target 
%, and the assumed $\lambda$ value from pQCD analysis~\cite{PhysRevD.99.014032}, 
were also studied. 
These were found to be small compared to the major sources discussed above.

%% file: res_discuss.tex
%\begin{figure}
%    \centering
%    \includegraphics[width=.24\textwidth]{mu_LH2_box.pdf}~
%    \includegraphics[width=.24\textwidth]{nu_LH2_box_all.pdf}\\
%    \includegraphics[width=.24\textwidth]{mu_LD2_box.pdf}~
%    \includegraphics[width=.24\textwidth]{nu_LD2_box_all.pdf}
%    \caption{Parameters of the angular distributions $\mu$ (left) and $\nu$ (right) obtained from LH$_2$ target (top) and LD$_2$ target (bottom) events (the red points and systematic uncertainty bands). The magenta triangle points show the E866 results~\cite{E866pd,E866pp}, the blue lines show the pQCD calculations~\cite{PhysRevD.99.014032}, the green triangle points show the NA10 results~\cite{NA10}, and blank circle points show the E615 results~\cite{E615}.}
%    \label{fig:e906_e866_pt}
%\end{figure}
\begin{figure}
    \centering
    \includegraphics[width=.5\textwidth]{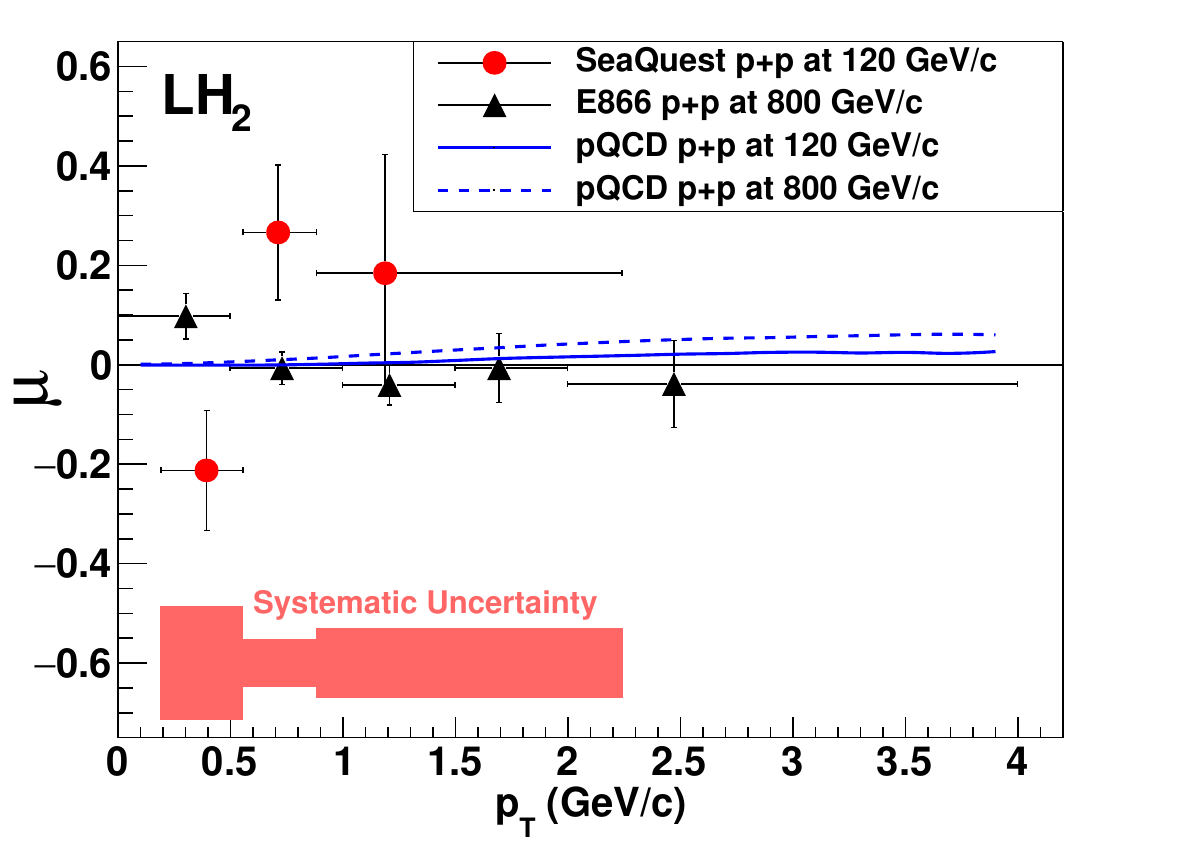}\\
    \includegraphics[width=.5\textwidth]{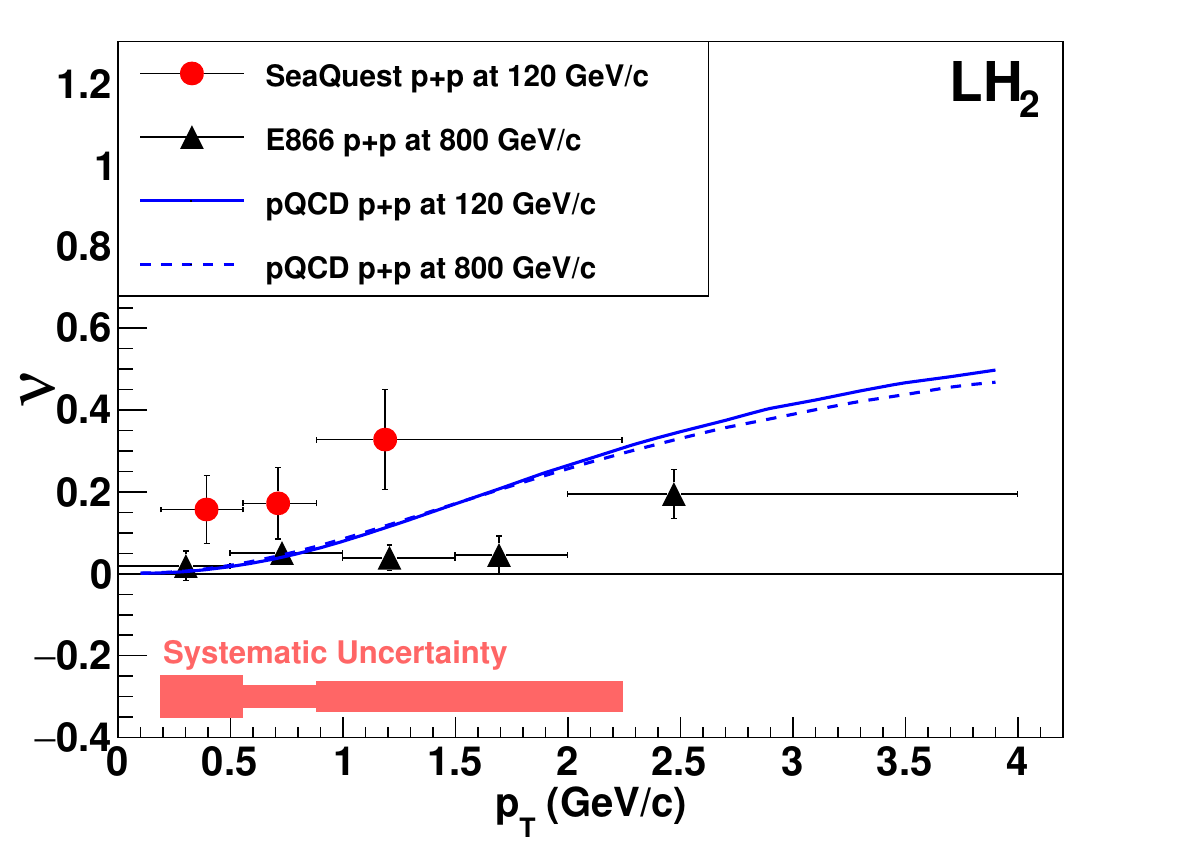}
    \caption{Parameters of the angular distributions $\mu$ (top) and $\nu$ (bottom) obtained from LH$_2$ target events (the red points and systematic uncertainty bands). The black triangle points show the E866 results~\cite{E866pp} and the blue lines show the pQCD calculations~\cite{PhysRevD.99.014032}}%, the green triangle points show the NA10 results~\cite{NA10}, and open circle points show the E615 results~\cite{E615_2}.}
    \label{fig:e906_e866_pt_pp}
\end{figure}
\begin{figure}
    \centering
    \includegraphics[width=.5\textwidth]{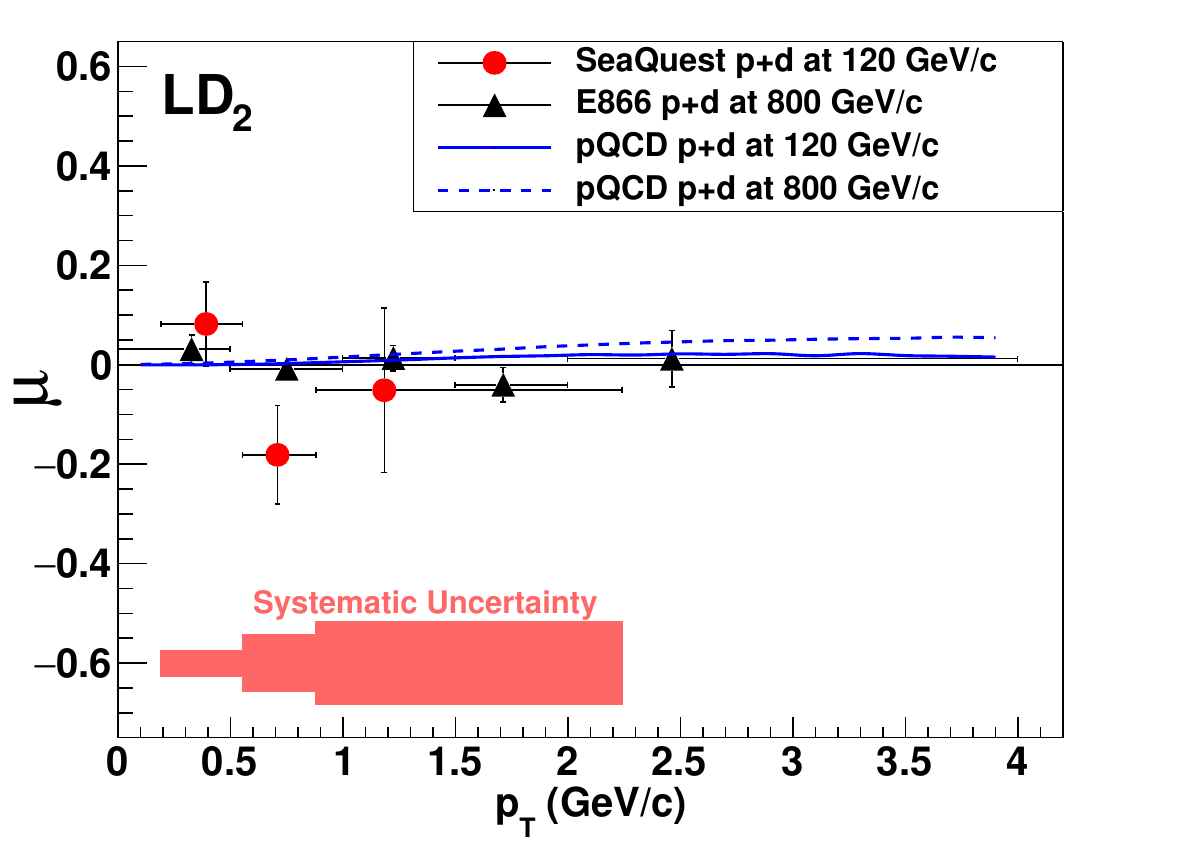}\\
    \includegraphics[width=.5\textwidth]{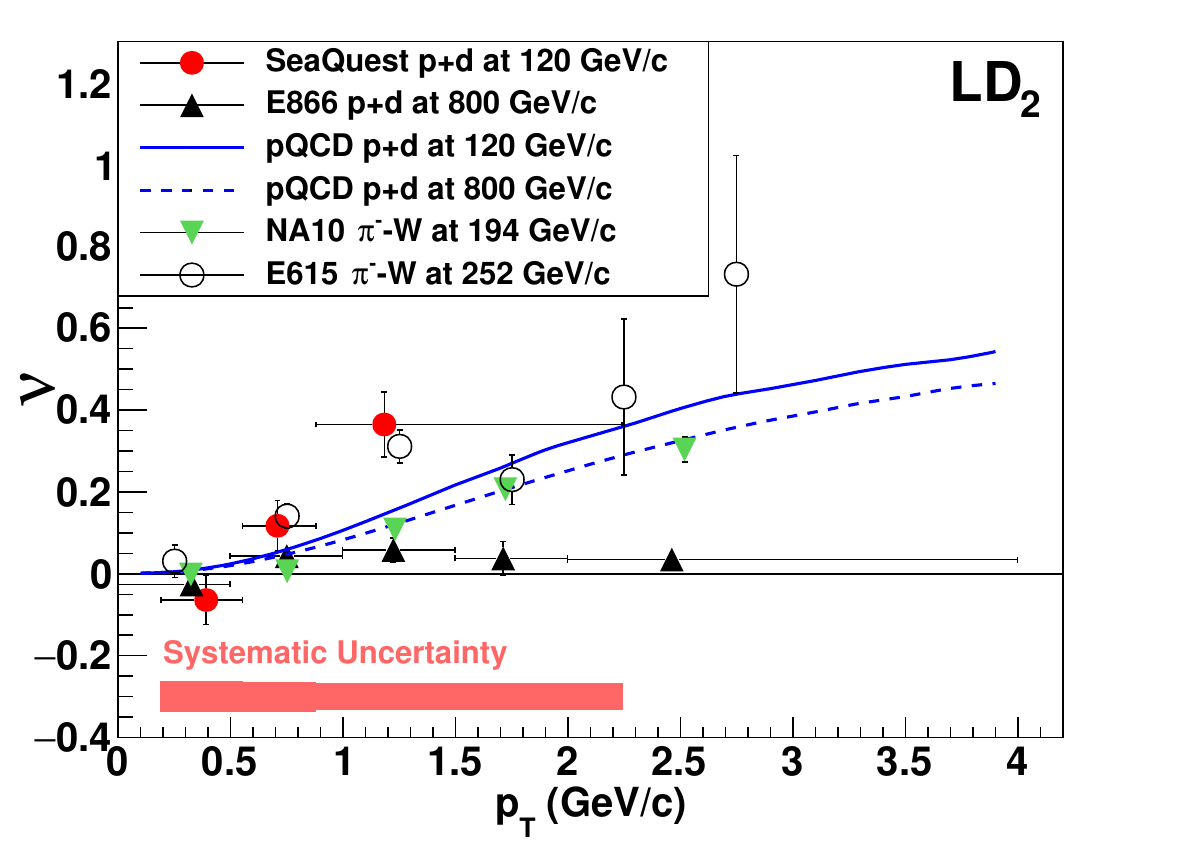}
    \caption{Parameters of the angular distributions $\mu$ (top) and $\nu$ (bottom) obtained from LD$_2$ target events (the red points and systematic uncertainty bands). The black triangle points show the E866 results~\cite{E866pd}, the blue lines show the pQCD calculations~\cite{PhysRevD.99.014032}, the green triangle points show the NA10 results~\cite{NA10}, and open circle points show the E615 results~\cite{E615_2}.}
    \label{fig:e906_e866_pt_pd}
\end{figure}
\begin{table}
    \centering
    \begin{tabular}{c|ccc}
        \hline
        \hline
        \multicolumn{4}{c}{LH$_2$}\\
        \hline
        $p_T$ bin (GeV/$c$) & 0.19-0.55 & 0.55-0.88 & 0.88-2.24 \\
        $\langle p_T\rangle$ (GeV/$c$)& 0.39 & 0.71 & 1.19 \\
        $\langle m_{\mu\mu}\rangle$ (GeV/$c^2$)& $5.2$ & $5.2$ & $5.3$ \\
        $\langle x_1\rangle$ & $0.56$ & $0.56$ & $0.56$ \\
        $\langle x_2\rangle$ & $0.23$ & $0.23$ & $0.24$ \\
        $\langle x_F\rangle$ & $0.38$ & $0.38$ & $0.37$ \\        
        \hline
%        $\lambda$ & 0.98 (fixed) & 0.92 (fixed) & 0.78 (fixed) \\
        $\mu$ & $-0.21 \pm 0.12 $ & $0.27  \pm 0.14 $ & $0.18 \pm 0.24$ \\
        & $\pm 0.11$ & $\pm0.05$ & $\pm 0.07$\\
        $\nu$ & $0.16 \pm 0.08$ & $0.17 \pm 0.09$ & $0.33 \pm 0.12$ \\
        & $\pm 0.05$ & $\pm 0.03$ & $\pm 0.04$\\
        \hline
%    \end{tabular}
%    \caption{Angular coefficients (LH$_2$ target)}
%    \label{tab:results_LH2}
%\end{table}
%\begin{table}
%    \centering
%    \begin{tabular}{c|ccc}
        \hline
        \multicolumn{4}{c}{LD$_2$}\\
        \hline
        $p_T$ bin (GeV/$c$) & 0.19-0.55 & 0.55-0.88 & 0.88-2.24 \\
        $\langle p_T\rangle$ (GeV/$c$) & 0.39 & 0.71 & 1.19 \\
        $\langle m_{\mu\mu}\rangle$ (GeV/$c^2$) & $5.2$ & $5.2$ & $5.3$ \\
        $\langle x_1\rangle$ & $0.57$ & $0.57$ & $0.57$ \\
        $\langle x_2\rangle$ & $0.22$ & $0.23$ & $0.24$ \\
        $\langle x_F\rangle$ & $0.39$ & $0.39$ & $0.38$ \\        
        \hline
%        $\lambda$ & 0.97 (fixed) & 0.89 (fixed) & 0.71 (fixed) \\
        $\mu$ & $0.08\pm0.08$ & $-0.18\pm 0.10$ & $-0.05\pm 0.17$ \\

        & $\pm 0.03$ & $\pm 0.06$ & $\pm 0.08$\\
            
        $\nu$ & $-0.06\pm0.06$ & $0.12\pm0.06$ & $0.36\pm0.08$ \\

        & $\pm 0.04$ & $\pm 0.04$ & $\pm 0.03$\\
        
        \hline
        \hline
    \end{tabular}
    \caption{\normalsize Parameters of the angular distributions obtained from LH$_2$ (top) and LD$_2$ (bottom) targets. The average kinematic variables in each $p_T$ bin are also shown. Both statistical (first error) and systematic (second error) uncertainties are shown.}
    \label{tab:results}
\end{table}

\begin{table}
    \centering
    \vspace{1em}
    \begin{tabular}{c||c|c|c}
        \hline
        & $\langle\lambda\rangle$ & $\langle\mu\rangle$ & $\langle\nu\rangle$ \\
        \hline
        E906 (SeaQuest) & & &  \\ 
        $p+p$ & N/A & $0.03\pm0.09$ & $0.20\pm0.05$ \\
        120 GeV/$c$ & & & \\
        \hline
        E906 (SeaQuest) & & & \\
        $p+d$ & N/A & $-0.07\pm0.06$ & $0.14\pm 0.04$ \\
        120 GeV/$c$ & & & \\
        \hline
        E866 & & & \\
        $p+p$ & & & \\ 
        800 GeV/$c$~\cite{E866pp} & $0.85\pm0.10$ & $-0.026\pm0.019$ & $0.040\pm0.015$ \\
        $\langle x_1\rangle=0.435$ \cite{personal} & & & \\
        $\langle x_2\rangle=0.074$ & & & \\
        \hline
        E866 & & & \\
        $p+d$ & & & \\
        800 GeV/$c$~\cite{E866pd} & $1.07\pm0.07$ & $0.003\pm0.013$ & $0.027\pm0.010$ \\
        $\langle x_1\rangle=0.433$ \cite{personal} & & & \\
        $\langle x_2\rangle=0.075$ & & & \\       
        \hline
        NA10 & & & \\
        $\pi^-+W$ & $0.83\pm0.04$ & $0.008\pm0.010$ & $0.091\pm0.009$ \\
        194 GeV/$c$~\cite{NA10} & & &\\
        \hline
        E615 & & & \\
        $\pi^-+W$ & $1.17\pm0.04$ & $0.09\pm0.02$ & $0.169\pm0.019$ \\
        252 GeV/$c$~\cite{E615_1} & & &\\
        \hline
    \end{tabular}
    \caption{Angular parameters averaged over the measured ranges.}
    \label{tab:comp}
\end{table}
Figure~\ref{fig:e906_e866_pt_pp}, Fig.~\ref{fig:e906_e866_pt_pd} and Tab.~\ref{tab:results} present the angular distribution parameters $\mu$ and $\nu$ as functions of $p_T$, extracted from the LH$_2$ and LD$_2$ data. 
Table~\ref{tab:comp} presents the averaged parameters over the measured kinematic range, compared with results from previous experiments. 
%The $p_T$ dependence of the parameters from earlier studies is shown in Fig.~\ref{fig:e906_e866_pt}.
The $\mu$ parameters are consistent with zero, in agreement with results from the E866 experiment and pQCD calculations \cite{PhysRevD.99.014032}. 
For the $\nu$ parameter, nonzero values are observed at large $p_T$, a trend similar to that reported in pion-induced Drell-Yan experiments (NA10 and E615). 
This behavior contrasts with the small $\nu$ values reported by E866, a proton-induced Drell-Yan experiment.
Since the $\lambda$ values were not extracted in this analysis, the Lam-Tung relation~\cite{Lam:1978pu}, $1 - \lambda = 2\nu$, which holds at next-to-leading order but is violated at next-to-next-to-leading order, was not tested in this analysis.

% The pQCD calculations predict large $\nu$ values at high $p_T$~\cite{PhysRevD.99.014032}. 
% The results from pion-induced Drell--Yan experiments show slight deviations from these predictions~\cite{PhysRevD.99.014032}.
% In contrast, the E866 experiment observed values near 0, much smaller than pQCD predictions at high $p_T$~\cite{PhysRevD.99.014032}, 
% which suggests suppression due to nonperturbative QCD contributions.
% The E906 results reveal deviations larger than those predicted by pQCD, providing a contrasting perspective.
% For the $p+p$ Drell--Yan process, the p-value of 3.5\% (6.5\% with systematic uncertainty added) indicates a mild but statistically significant positive deviation, implying potential contributions from nonperturbative QCD.
% The $p+d$ Drell--Yan results show a more pronounced deviation, with a p-value of 1.5\% (3.5\% with systematical uncertainty added), suggesting an enhanced role of nonperturbative QCD effects in deuterium targets.

The present analysis reveals a trend of increasing $\nu$ values with $p_T$ in both $p+p$ and $p+d$ collisions at 120 GeV/$c$, consistent with the behavior predicted by pQCD calculations \cite{PhysRevD.99.014032}. 
However, as shown in Figs.~\ref{fig:e906_e866_pt_pp} and \ref{fig:e906_e866_pt_pd} (bottom), the observed magnitudes significantly exceed the pQCD predictions. For the $p+p$ Drell-Yan process, the p-value of 3.5\% (6.5\% with systematic uncertainty included) indicates a statistically significant positive deviation, suggesting contributions from nonperturbative QCD effects. 
The $p+d$ Drell-Yan results exhibit an even more pronounced deviation, with a p-value of 1.5\% (3.5\% with systematic uncertainty included), pointing to an enhanced role of nonperturbative QCD in deuterium targets. 
Similarly, the pion-induced Drell-Yan experiments (NA10~\cite{NA10_2} and E615~\cite{E615_1,E615_2}) observed $\nu$ values exceeding pQCD predictions \cite{PhysRevD.99.014032}, with the significant $\mathrm{cos}(2\phi)$ modulation attributed to the Boer-Mulders mechanism. 
The E906 results align with this pattern of enhancements beyond pQCD expectations. 
In contrast, the Fermilab E866/NuSea experiment~\cite{E866pp, E866pd} measured $\nu$ values smaller than pQCD predictions at high $p_T$ \cite{PhysRevD.99.014032}, yielding a distinctly different conclusion from other fixed-target Drell-Yan measurements.

%This work was performed by the SeaQuest Collaboration, whose work was supported in part by the US Department of Energy under Grant No. DE-AC02-06CH11357 and DE-FG02-07ER41528; 
%the US National Science Foundation under Grants No. PHY 2309922, PHY 2013002, PHY 2110229, PHY 2209348, and PHY 2514181;
%the National Science and Technology Council (NSTC) of Taiwan under Grant No.112-2112-M-008-041-MY3;
%the JSPS (Japan) KAKENHI through Grants No.21244028, No.25247037, No.25800133, No.18H03694, No.20K04000, No.22H01244, No.23K22515, No.25K07341.

%We thank G. T. Garvey, C.N. Brown, and N. Makins for contributions to the early stages of this experiment.
%We also thank the Fermilab Accelerator Division and Particle Physics Division for their support of this experiment. 
We thank the late G. T. Garvey for contributions to the early stages of this experiment, 
C. N. Brown for contributions to 5 decades of dimuon experiments at Fermilab, 
and N. C. R. Makins for contributions to the execution of the experiment. 
We also thank the Fermilab Accelerator Division and Particle Physics Division for their support of this experiment. 
This work was performed by the SeaQuest Collaboration, whose work was supported in part by the U.S. Department of Energy, Office of Nuclear Physics under contract No. DE-AC02-06CH11357, DE-SC0013620, DE-FG02-07ER41528; 
the US National Science Foundation under Grants No. PHY 2013002, PHY 2110229, PHY 2111046, PHY 2209348, PHY 2309922, PHY 2514181; 
the JSPS (Japan) KAKENHI through Grants No. 21244028, No. 25247037, No. 25800133, No. 18H03694, No. 20K04000, No. 22H01244, No. 23K22515, No. 25K07341; 
and the National Science and Technology Council of Taiwan (R.O.C.). 
Fermilab is managed by FermiForward Discovery Group, LLC, acting under Contract No. 89243024CSC000002.